\documentclass[runningheads]{llncs}
\usepackage[T1]{fontenc}
\usepackage{graphicx,verbatim}
\usepackage{hyperref}
\usepackage{color}

\usepackage{amsmath} 
\usepackage{amssymb}
\usepackage{booktabs}
\usepackage[table]{xcolor}
\def\ourcolor{cyan!45}
\newcommand{\tabbl}{\midrule \rowcolor{lightgray!28}}
\definecolor{cyan}{cmyk}{.3,0,0,0}
\definecolor{hr}{gray}{0.65}
\def\ourcolor{cyan!45}
\definecolor{hg}{gray}{0.8}
\def\onedot{.}
\def\eg{{\em e.g}\onedot}

\def\modelname{APL}

\begin{document}
\title{Adaptive Prototype Learning for Multimodal Cancer Survival Analysis}
\author{Hong Liu\inst{*,1} \and
Haosen Yang\inst{*,2} \and
Federica Eduati\inst{1} \and
Josien P.W. Pluim\inst{1} \and
Mitko Veta\inst{1}
}

\authorrunning{H.Liu et al.}

\institute{Eindhoven University of Technology, Eindhoven, The Netherlands 
\and
University of Surrey, London, UK\\
*Joint main authors \email{\{hong.liu.00408,haosen.yang.6\}@gmail.com}}

\maketitle              

\begin{abstract}
Leveraging multimodal data, particularly the integration of whole-slide histology images (WSIs) and transcriptomic profiles, holds great promise for improving cancer survival prediction. However, excessive redundancy in multimodal data can degrade model performance.  
In this paper, we propose Adaptive Prototype Learning (\textbf{\modelname{}}), a novel and effective approach for multimodal cancer survival analysis. \modelname{} adaptively learns representative prototypes in a data-driven manner, reducing redundancy while preserving critical information. Our method employs two sets of learnable query vectors that serve as a bridge between high-dimensional representations and survival prediction, capturing task-relevant features. Additionally, we introduce a multimodal mixed self-attention mechanism to enable cross-modal interactions, further enhancing information fusion.  
Extensive experiments on five benchmark cancer datasets demonstrate the superiority of our approach over existing methods. The code is available at \url{https://github.com/HongLiuuuuu/APL}.


\end{abstract}

\section{Introduction}
Survival analysis, a cornerstone of patient prognostic modeling, aims to predict the time until an event of interest occurs (typically death), thereby improving therapeutic decision-making, optimizing patient care, and aiding in the identification of novel biomarkers associated with disease progression~\cite{Song2023AIPathology}.
Prognostication is a complex challenge influenced by diverse perspectives~\cite{Cancer}. Multimodal methods that integrate features from histology and genomics data can offer complementary insights, capturing subtle changes that may remain undetected within single-modality analyses~\cite{cmta,Cancer,mcat,motcat,survpath,pibd,mmp,ccl}.
Histology provides detailed phenotypic insights into cell types and the tumor microenvironment~\cite{bio1,bio2}. Genomics data, such as bulk transcriptomics~\cite{Acosta2022}, represents gene expression, revealing a comprehensive landscape of molecular information~\cite{Cancer,LIPKOVA2022,Steyaert2023,mmp}. Since these modalities capture distinct aspects of tumor biology, their integration enables a more holistic characterization of disease progression.

\begin{figure}[t]
    \centering
    \includegraphics[width=0.98\linewidth]{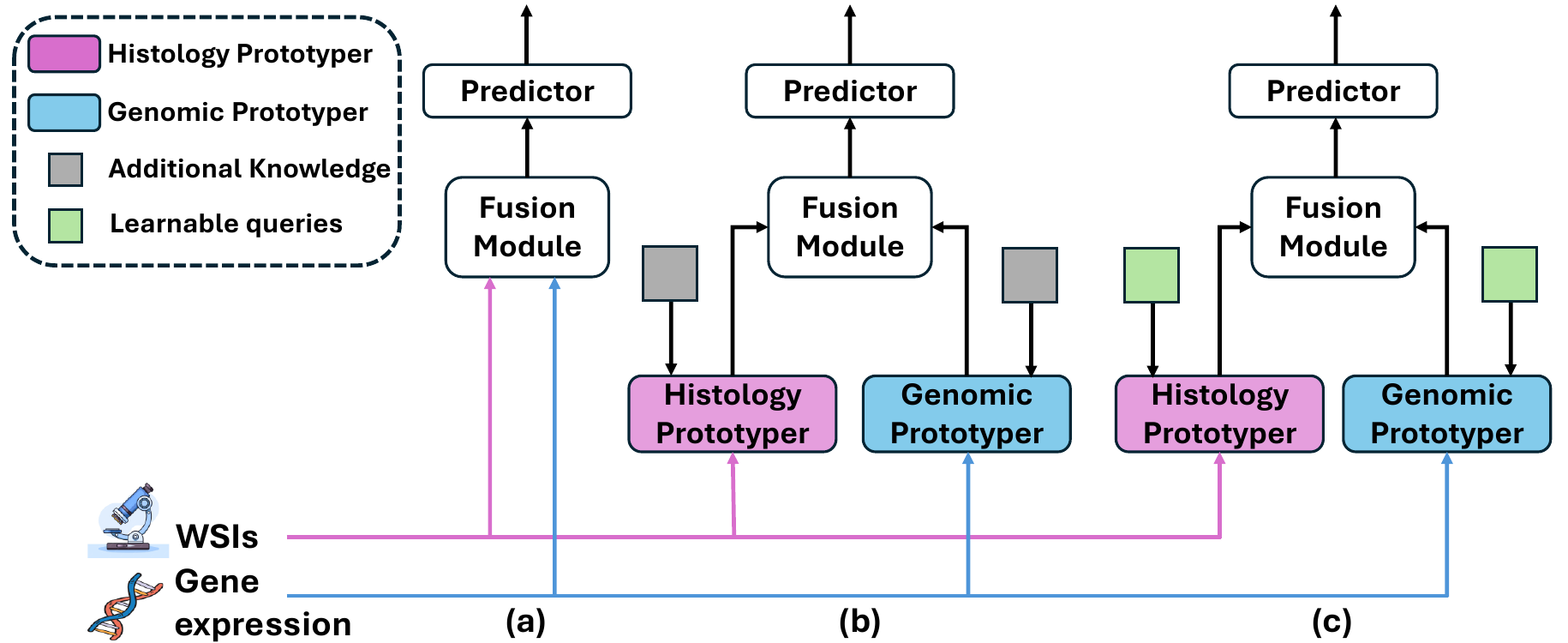}
    \caption{Illustration of typical multimodal cancer survival analysis architectures: (a) Directly fusing multimodal data through a fusion module, such as an attention mechanism (\eg, SurvPath~\cite{survpath}). (b) Reducing redundant tokens from cross-modal data using additional knowledge, such as predefined risk levels (\eg, PIBD~\cite{pibd}). (c) Our proposed approach adaptively learns task-relevant prototypes with learnable queries.}
    \label{fig1}
\end{figure}

In recent years, various multimodal methods~\cite{mcat,Cancer,cmta,motcat,survpath}  have combined these two modalities to enhance precision in risk stratification and optimize survival analysis (Figure~\ref{fig1}(a)). However, these works are hampered by the extensive number of histology and genomic tokens (\eg, patches of WSIs and pathways of gene expression), leading to information redundancy issue~\cite{HOSSEINI2024100357,pibd}. Several approaches~\cite{mmp,pibd,ccl}have been proposed to incorporate additional knowledge by clustering tokens into fixed categories, thereby reducing the number of cross-modal tokens (Figure~\ref{fig1}(b)) . We refer to these representative categories as \textbf{predefined prototypes}. For instance, MMP~\cite{mmp} groups all histology tokens into morphology-related categories, while PIBD~\cite{pibd} and CCL~\cite{ccl} cluster large token sets based on risk levels and censorship knowledge.
Although these methods significantly reduce the number of cross-modal tokens, they remain suboptimal in compacting extensive histology and genomic information. This limitation arises from their reliance on predefined prototypes based on morphology, risk levels, or censorship, restricting their flexibility in capturing dynamic data changes and emerging patterns. 

To address these limitations, we propose \textbf{\modelname}, a straightforward yet effective approach that adaptively learns representative prototypes in a data-driven manner (Figure~\ref{fig1}(c)). Our method begins by extracting unimodal representations of pathology and genomics data, following~\cite{survpath}.
To mitigate redundancy without relying on additional knowledge, we introduce two sets of learnable query vectors that interact with pathology and genomic features through cross-attention mechanisms. These queries extract compact representations, which serve as pathology and genomic prototypes, enabling the model to distill essential features from high-dimensional data.
The learnable queries act as a bridge between high-dimensional representations and survival analysis, capturing task-relevant features while minimizing redundancy without requiring additional knowledge.
To further enhance multimodal fusion, we employ a multimodal mixed self-attention mechanism on the combined set of histology and genomic prototypes, enabling the model to learn cross-modal interactions and improve information fusion.

We summarize the contributions as follows:  (1) We propose \modelname{}, a novel multimodal framework designed to mitigate information redundancy in cancer survival analysis.
(2) To achieve this, we introduce two sets of learnable queries for pathology and genomic prototypes, leveraging cross-attention to capture task-relevant features and multimodal mixed self-attention mechanism to model cross-modal interactions.
(3) Extensive evaluations on five benchmark cancer datasets demonstrate the effectiveness of \modelname.

\begin{figure}[t]
    \centering
    \includegraphics[width=0.98\linewidth]{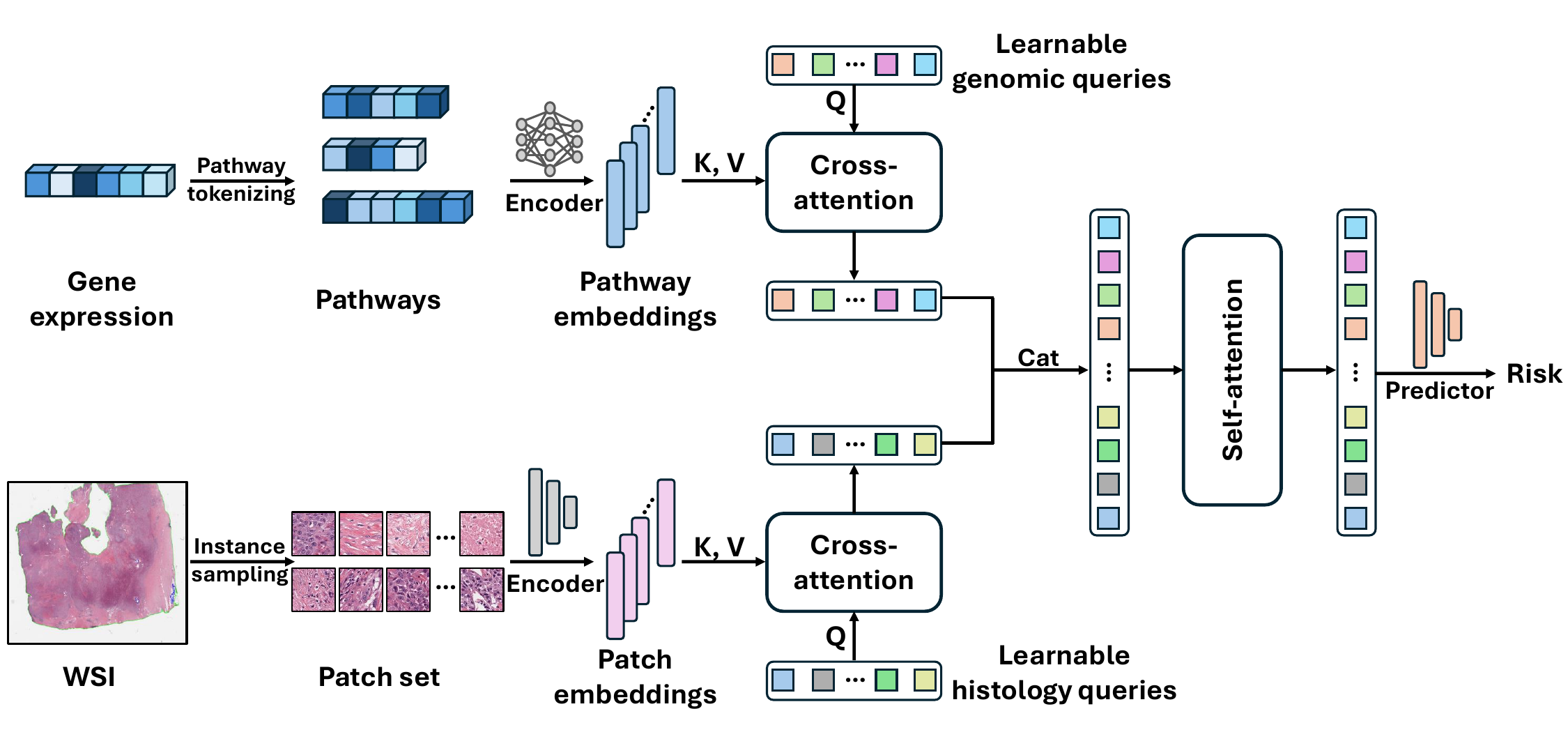}
    \caption{Overview of \textbf{\modelname{}}. Gene expression is first tokenized into biological pathways, and pathway embeddings are extracted using a feature extractor (SNN~\cite{snn}). Similarly, WSIs are processed into patch embeddings using a pre-trained feature extractor. Next, an adaptive prototyping module employs two sets of learnable queries to extract compact information from high-dimensional representations via cross-attention. These learned prototypes are then fused using a multimodal mixed self-attention mechanism, facilitating cross-modal interactions and enhancing information integration. Finally, the model predicts survival risk based on the refined prototypes.}
    \label{fig2}
\end{figure}
\section{Methods}
We introduce the \textbf{\modelname{}} framework, which learns histology and genomics prototypes in a data-driven manner for survival prediction. 
First, we describe adaptive prototyping mechanism with learnable queries in Section~\ref{sec:apl}. Then, we outline the multimodal fusion mechanism and describe survival prediction in Section~\ref{sec:fusion}.

\subsection{Adaptive prototyping mechanism}\label{sec:apl}
The adaptive prototyping mechanism aims to reduce redundancy in extracted unimodal features using learnable queries. The details are outlined below.
\paragraph{Unimodal feature extraction.} 
We start by extracting unimodal features from histology and genomic data. WSIs capture detailed tissue phenotypes, offering critical insights for cancer prognosis prediction. To ensure a focus on biologically relevant information, we first identify tissue regions, excluding background areas that lack diagnostic significance.
Because of the extremely high resolution of WSIs, we divide the identified tissue regions into a set of \(N_H\) non-overlapping patches at \(20\times\) magnification, denoted as 
\(
H = \{h_1, ..., h_{N_H}\}, N_H > 10^4
\).
Due to the large number of patches per WSI, storing and processing them all at once is impractical. To overcome this, we extract patch embeddings before training. Specifically, we use a pre-trained feature extractor \( f(\cdot) \) to map each patch \( h_i \) into a patch embedding $x^{(H)}_i = f(h_i)$.
We apply a learnable linear transformation, yielding the final patch features $X^{(H)}$.
This process efficiently represents WSIs while preserving biologically meaningful information for downstream survival prediction tasks.

Bulk transcriptomics captures gene expression patterns that reflect the molecular state of a tumor, including its aggressiveness and response to treatment. These molecular signatures provide valuable prognostic information, making transcriptomics a powerful tool for predicting patient survival. Following~\cite{survpath}, we construct pathways by grouping genes with known interactions relevant to specific cellular processes. To standardize variable-length pathways, we encode them into fixed-length genomic embeddings \( X^{(G)} \) using self-normalizing neural networks (SNN)~\cite{snn}.

\paragraph{Learnable histology and genomics prototypes.}
While patch and pathway features provide valuable insights for survival analysis, their sheer volume limits the effective application of attention mechanisms to capture comprehensive information.  
To address this, we introduce learnable histology and genomic queries, denoted as \( Q^{(H)} \) and \( Q^{(G)} \), to learn histology and genomic prototypes. These queries interact with patch and pathway features, extracting compact representations from high-dimensional data via a cross-attention mechanism. In this mechanism, learnable histology tokens \( Q^{(H)} \) serve as queries, while histology patch embeddings \( X^{(H)} \) function as both keys and values. The compact histology feature can be obtained with:
\begin{equation} 
Q'^{(H)}  = \text{Softmax} \left( \frac{F_{q}(Q^{(H)}) F_{k}(X^{(H)})}{\sqrt{C}} \right) F_{v}(X^{(H)})
\end{equation}
where \( F_q \), \( F_k \), and \( F_v \) denote linear projection functions for queries, keys and values, and \( C \) represents the dimensionality of the projected features. Similarly, the compact genomic feature \( Q'^{(G)} \) can be obtained using the same approach.
In our work, learnable queries act as a bridge between high-dimensional representations and survival analysis, capturing task-relevant prototypes while reducing redundancy.

\subsection{Multimodal fusion and prediction}\label{sec:fusion}
We propose a multimodal mixed self-attention mechanism to enhance dense interactions between the compact histology and genomic information.
To achieve this, we define a multimodal sequence by concatenating the compact histology and genomic feature, denoted as:
\[
M^{(HG)} = Q'^{(H)} \| Q'^{(G)}
\]
where \( Q'^{(H)} \) and \( Q'^{(G)} \) represent the compact histology and genomic feature, respectively.   
Followed by a self-attention~\cite{attn}, efficient connections are established between both modalities. Then we have the fused prototypes 
\begin{equation} 
M'^{(HG)}  = \text{Softmax} \left( \frac{F_q^{1}(M^{(HG)}) F_k^{1}(M^{(HG)})}{\sqrt{C}} \right) F_v^{1}(M^{(HG)})
\end{equation}
where \( F_q^{1} \), \( F_k^{1} \), and \( F_v^{1} \) denote linear projection functions. 
Rather than decomposing the multimodal transformer attention into four intra- and cross-modality parts~\cite{survpath}, we directly employ self-attention to measure and aggregate interactions among all multimodal prototypes. This approach leverages the advantage of a reduced number of prototypes, ensuring efficient and effective information fusion. We average the fused feature and pass them through a predictor to obtain the logit, representing the probability of death within a given time interval.

Our objective is to predict patient survival, formulating survival risk prediction as a classification task, following previous research~\cite{survpath,pibd}. To train the model, we use the negative log-likelihood (NLL) loss~\cite{nll} to supervise survival prediction.

\section{Experiments}
\paragraph{Datasets.}
We performed extensive experiments using five public cancer datasets from The Cancer Genome Atlas (TCGA)\setcounter{footnote}{0}\footnote{\url{https://portal.gdc.cancer.gov/}}: Breast Invasive Carcinoma (BRCA, n=869), Bladder Urothelial Carcinoma (BLCA, n=359), Head and Neck Squamous Cell Carcinoma (HNSC, n=392), Colon and Rectum Adenocarcinoma (COADREAD, n=296), and Stomach Adenocarcinoma (STAD, n=317). 
We trained models to predict disease-specific survival (DSS)~\cite{survpath}, which more accurately reflects the patient's disease status compared to overall survival.
For histology data, we extracted non-overlapping 224 × 224 patches at 20× magnification. 
For genomic data, raw transcriptomics were obtained from the Xena database~\cite{Goldman2020Xena}, along with DSS labels. 331 human biological pathways were collected, represented as transcriptomics sets with specific molecular interactions, sourced from the Human Molecular Signatures Database (MSigDB) - Hallmarks~\cite{LIBERZON2015417,0506580102} (50 pathways from 4,241 genes) and Reactome~\cite{reactome} (281 pathways from 1577 genes), ensuring at least 90\% of transcriptomics were accessible.

\paragraph{Evaluation Metrics.}
To reduce potential batch artifacts, we use 5-fold cross-validation for each dataset. Model performance is evaluated using the concordance index (C-index)~\cite{cindex} and its standard deviation (std), which measures the accuracy of ranking patients based on their survival months and predicted risk.

\paragraph{Implementation details.}
The proposed algorithm is implemented in Python with Pytorch library and runs on a single NVIDIA A100 GPU. UNI~\cite{uni}, a
DINOv2-based ViT-Large~\cite{dinov2} model pretrained on \(1 \times 10^{8}\) patches sampled from
\(1 \times 10^{5}\) WSIs collected at Mass General Brigham, is used as the feature extractor to get 1024-dimensional embeddings. We further use an MLP with a 512-dimensional hidden layer as the latent vector encoder to embed patch features into a fixed dimension of 256. Meanwhile, the feature extractors of pathways are SNNs following the settings in works~\cite{survpath,mmp,pibd,ccl}.
All models are trained with a \(5 \times 10^{-4}\) learning rate with \(1 \times 10^{-3}\) weight decay for 50 epochs, AdamW optimizer~\cite{adam} and the batch size is set to 32. we set the number of histology and genomic learnable queries to 300 and 128. 

\begin{table}[t]
    \centering
    \caption{Comparison of \modelname{} and baseline methods for disease-specific patient survival prediction, measured by the C-Index. The best performance is highlighted in bold. * Indicates prototype-based methods.}
    \renewcommand{\arraystretch}{1.2}
    \resizebox{\textwidth}{!}{
    \begin{tabular}{l  c c c c c c}
        \toprule
        Model & BRCA & BLCA & COADREAD  & HNSC & STAD & Avg. \\
        \tabbl
        \multicolumn{7}{l}{\small{\it  Genomic}} \\
        MLP & $0.598 \pm 0.063$ & $0.501 \pm 0.071$ & $0.709 \pm 0.158$ & $0.512 \pm 0.057$ & $0.479 \pm 0.052$ & $0.560$ \\
        SNN & $0.639 \pm 0.067$ & $0.584 \pm 0.067$ & $0.732 \pm 0.134$ & $0.567 \pm 0.055$ & $0.557 \pm 0.051$ & $0.616$ \\
        \tabbl
        \multicolumn{7}{l}{\small{\it  Histology}} \\
        ABMIL & $0.642 \pm 0.065$ & $0.612 \pm 0.065$ & $0.702 \pm 0.148$ & $0.619 \pm 0.048$ & $0.608 \pm 0.054$ & $0.636$ \\
        AMISL & $0.613 \pm 0.046$ & $0.601 \pm 0.053$ & $0.694 \pm 0.123$ & $0.602 \pm 0.054$ & $0.559 \pm 0.032$ & $0.614$ \\
        \tabbl
        \multicolumn{7}{l}{\small{\it  Multimodal}} \\
        Porpoise & $0.642 \pm 0.043$ & $0.619 \pm 0.056$ & $0.702 \pm 0.143$ & $0.631 \pm 0.042$ & $0.639 \pm 0.075$ & $0.646$ \\
        MCAT & $0.713 \pm 0.033$ & $0.632 \pm 0.066$ & $0.715 \pm 0.158$ & $0.635 \pm 0.098$ & $0.668 \pm 0.087$ & $0.673$ \\
        MOTCat & $0.712 \pm 0.042$ & $0.641 \pm 0.067$ & $0.728 \pm 0.134$ & $0.641 \pm 0.064$ & $0.658 \pm 0.066$ & $0.676$ \\
        SurvPath & $0.723 \pm 0.045$ & $0.642 \pm 0.054$ & $0.726 \pm 0.161$ & $0.646 \pm 0.057$ & $0.649 \pm 0.051$ & $0.677$ \\
        PIBD* & $0.716 \pm 0.026$ & $0.650 \pm 0.067$ & $0.734 \pm 0.153$ & $0.642 \pm 0.054$ & $0.656 \pm 0.051$ & $0.680$ \\
        MMP* & $0.746 \pm 0.064$ & $0.660 \pm 0.050$ & $0.741 \pm 0.168$ & $0.641 \pm 0.046$ & $0.640 \pm 0.037$ & $0.686$ \\
        CCL* & $0.772 \pm 0.066$ & $0.662 \pm 0.055$ & $0.758 \pm 0.118$ & $0.629 \pm 0.047$ & $0.632 \pm 0.053$ & $0.690$ \\
        \rowcolor{\ourcolor} \textbf{APL}* & $\textbf{0.794} \pm \textbf{0.062}$ & $ \textbf{0.677} \pm \textbf{0.060}$ & $\textbf{0.812} \pm \textbf{0.115}$ & $\textbf{0.653} \pm \textbf{0.045}$ & $\textbf{0.686} \pm \textbf{0.053}$ & $\textbf{0.724}$ \\
        \bottomrule
    \end{tabular}
    \label{tab:main}
    }
\end{table}

\paragraph{Comparison to state of the art.}
We evaluate our method against three groups of SOTA approaches:
(1) Unimodal Methods: For genomic data, we use MLP~\cite{Haykin1998NeuralNetworks} and SNN~\cite{snn} as baselines. For histology, we compare against ABMIL~\cite{abmil} and AMISL~\cite{amisl}.
(2) Multimodal Methods: We benchmark our approach against four leading multimodal models: Porpoise~\cite{Cancer}, MCAT~\cite{mcat}, MOTCat~\cite{motcat}, and SurvPath~\cite{survpath}.
(3) Prototype-based Methods: We further compare our method with three prototype-based multimodal approaches: PIBD~\cite{pibd}, MMP~\cite{mmp}, and CCL~\cite{ccl}.

The results, summarized in Table~\ref{tab:main}, indicate that \modelname{} consistently outperforms all other methods across five cancer datasets, achieving the highest average C-index of 0.724. Compared to unimodal approaches, \modelname{} surpasses the best-performing genomic and histology models (SNN: 0.616, ABMIL: 0.636), respectively, highlighting the advantage of integrating multimodal information and the significance of effectively mitigating information redundancy.

Among multimodal methods, \modelname{} achieves the highest performance across all four benchmarks, surpassing the second-best method (CCL: 0.690) by 3.4 percentage points in average C-index. Furthermore, within the prototype-based multimodal group, \modelname{} demonstrates clear superiority, achieving performance gains ranging from 0.7 to 5.4 percentage points compared to PIBD, MMP, and CCL. By leveraging learnable queries as an intermediary between high-dimensional representations, \modelname{} effectively mitigates information redundancy and enhances survival prediction, confirming its effectiveness in multimodal cancer analysis.
\begin{table}[t]
    \centering
     \caption{Ablation study on different configurations of \modelname.}
    \begin{tabular}{c c c|c c c c}
        \hline
        Hist. & Geno. & Self-attn. & BRCA & BLCA & COADREAD & Avg. \\
        \hline
          &   &   & $0.724 \pm 0.0612 $& $0.651 \pm 0.034 $ & $0.754 \pm 0.146$ & 0.709  \\
        $\checkmark$ &   &   & $0.745 \pm 0.0787 $  & $0.660 \pm 0.051 $ & $0.779 \pm 0.144 $ & 0.734  \\
        $\checkmark$ & $\checkmark$ &   & $0.761 \pm 0.0624 $ & $0.673 \pm 0.032 $ & $0.801  \pm 0.1426 $& 0.745 \\
        $\checkmark$ & $\checkmark$ & $\checkmark$ &$\textbf{0.794} \pm \textbf{0.062}$ & $\textbf{0.677} \pm \textbf{0.060}$ & $\textbf{0.812} \pm \textbf{0.115}$ & $\textbf{0.761}$ \\
        \hline
    \end{tabular}
   
    \label{tab:ablation}
\end{table}

\paragraph{Ablation study.}
We conduct ablation studies on BRCA, BLCA, and COADREAD datasets to evaluate the impact of three key components of \modelname{}: learned histology prototypes, learned genomics prototypes, and the multimodal mixed self-attention mechanism.  

We start with a simple baseline that directly concatenates the extracted histology and genomic features, followed by a predictor for survival prediction. Next, we integrate learned histology prototypes (Hist.) into the baseline, effectively reducing redundancy in histology features and improving the average metric from 0.709 to 0.734, demonstrating its importance.  

\begin{figure}[t]
    \centering
    \includegraphics[width=0.98\linewidth]{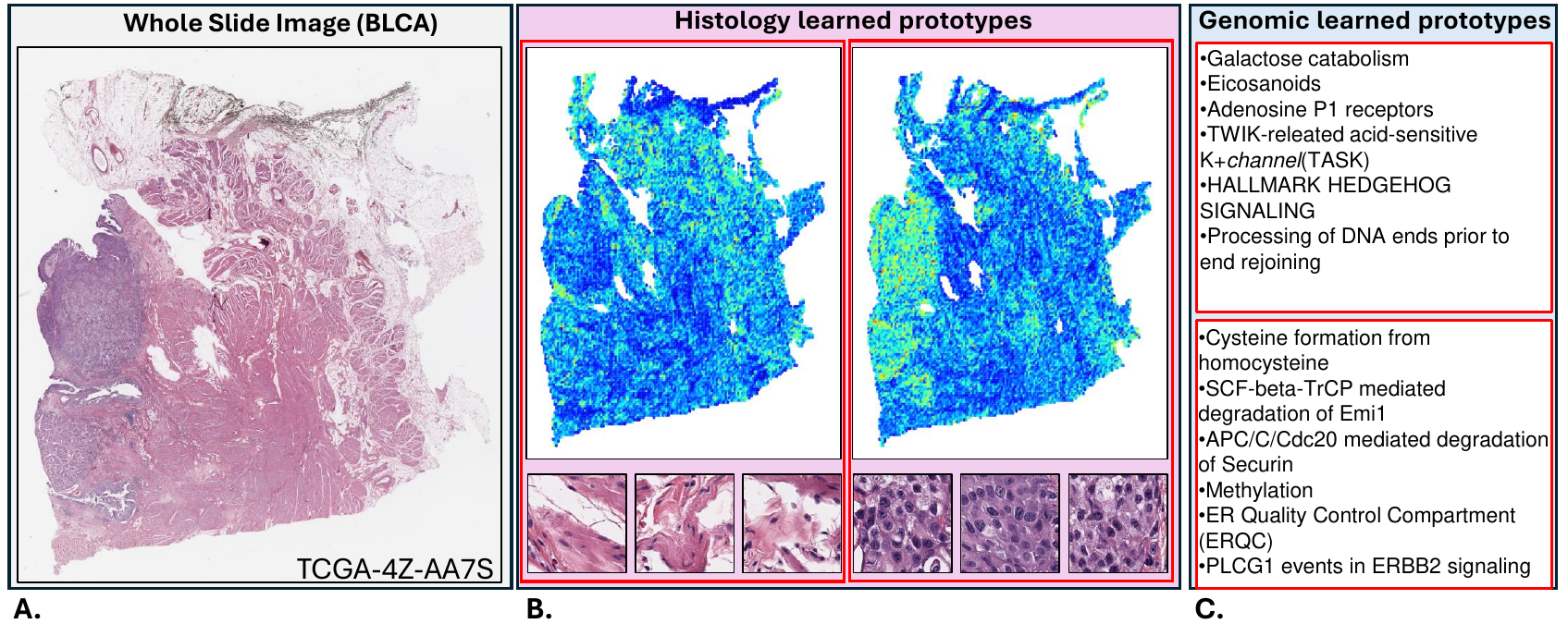}
    \caption{Visualization of \modelname{}'s behavior, including cross-attention maps and learned prototypes for histology and genomics. (A) WSI of a BLCA patient. (B) Top: Cross-attention maps of two randomly selected histology prototypes, where brighter regions indicate higher relevance. Bottom: The top three most representative patches corresponding to each learned prototype. (C) The top six pathways associated with two randomly selected genomic prototypes. Each histology and genomic prototype is highlighted in a red box.
   }
    \label{fig3}
\end{figure}

Incorporating learned genomics prototypes (Geno.) further enhances performance consistently across all three datasets. Finally, we introduce multimodal mixed self-attention mechanism (Self attn.), enabling the model to learn iterative cross-modal interactions and improve multimodal information fusion, achieving the best overall performance.
\paragraph{Model behavior visualization.}
To gain an intuitive understanding of \modelname{}'s impact, we analyze its behavior by examining the cross-attention maps and learned prototypes for histology and genomics on a BLCA case, as shown in Figure~\ref{fig3}(A).

For histology, we randomly select two learned queries as histology prototypes, using patch features as keys and values. The attention maps in the top of Figure~\ref{fig3}(B) illustrate how these prototypes are distributed across the WSI, where brighter regions indicate higher relevance to the current prototype. To further interpret these prototypes, we visualize the three most representative patches corresponding to each learned histology prototype (bottom of Figure~\ref{fig3}(B)). The distinct patterns captured by different prototypes demonstrate their ability to encode diverse histological features.

Similarly, for genomics, we select the top six pathways associated with two randomly chosen learned genomic prototypes, as shown in Figure~\ref{fig3}(C). The results highlight that different genomic prototypes are linked to distinct biological pathways, further validating their ability to capture meaningful genomic variations.

\section{Conclusion and Limitations}
In this paper, we introduced \modelname{}, a novel approach for multimodal cancer survival analysis. By adaptively learning representative prototypes, \modelname{} reduces redundancy while preserving critical information. It employs learnable query vectors to capture task-relevant features and a multimodal mixed self-attention mechanism to enhance cross-modal interactions.  
Experiments on five benchmark cancer datasets confirm the superiority of \modelname{} over existing methods, highlighting its effectiveness in improving cancer survival prediction.

While our method reduces the number of tokens, it requires a fixed number of learnable queries for both histology and genomics data across all datasets. This constraint may not be optimal, and exploring the dynamic number of queries remains an avenue for future work.

\bibliographystyle{splncs04}
\bibliography{mybibliography}

\end{document}